\documentclass[aps,prl,twocolumn,amsmath,amssymb,showpacs]{revtex4-1}
\usepackage{graphicx}
\usepackage{dcolumn}
\usepackage{bm}

\begin{document}

\title{
Delayed collapses of BECs in relation to AdS gravity}
\author {Anxo F. Biasi$^{1}$, Javier Mas$^{1}$ and Angel Paredes$^{2}$}

\affiliation{
 $^{1}$Departamento de F\'\i sica de Part\'\i culas, Universidade de Santiago de Compostela
 and Instituto Galego de F\'\i sica de Altas Enerx\'\i as (IGFAE), E-15782, Santiago de Compostela, Spain.\\
$^{2}$ Departamento de F\'\i sica Aplicada,
Universidade de Vigo, As Lagoas s/n, Ourense, E-32004 Spain.}

\begin{abstract}

We numerically investigate  spherically symmetric collapses in the Gross-Pitaevskii equation 
with attractive nonlinearity in a harmonic  potential.  Even below threshold for direct collapse, 
the wave function bounces off from the origin and may eventually become singular after a number of 
oscillations  in the trapping potential. This is reminiscent of 
 the evolution of Einstein gravity sourced by a scalar field
in Anti-de Sitter space where collapse corresponds to black hole formation. We carefully examine the 
long time evolution of the wave function for continuous families of initial states in order to sharpen out this 
qualitative
coincidence which may bring new
insights in both directions.  On one hand, we comment on possible implications for the so-called Bosenova collapses in
cold atom Bose-Einstein condensates. On the other hand, Gross-Pitaevskii 
  provides a toy model to 
study the relevance of either the  resonance conditions or the nonlinearity for the problem 
of Anti-de Sitter instability. 

\end{abstract}


\maketitle

\section{I. Introduction}
 
The nonlinear Schr\"odinger equation (NLSE), sometimes called
Gross-Pitaevskii equation (GPE) is a paradigmatic model for nonlinear systems.
Different versions have been used in diverse contexts including, {\it e.g.}
optics \cite{agrawal2007nonlinear,malomed2005spatiotemporal}, 
plasma physics \cite{zakharov2012kolmogorov} and even in cosmology 
\cite{schive2014cosmic,paredes2016interference}. 
It  describes with precision most of the physics of Bose-Einstein condensed (BEC) dilute
gases \cite{dalfovo1999theory,carretero2008nonlinear}. 
With an attractive nonlinearity, the wavefunction can self-focus and  collapse,
diverging at some point of space in finite time \cite{sulem2007nonlinear}.
This can happen in cold atom BECs when the  scattering length $a$ controlling
the effective atom-atom interaction is negative 
\cite{PhysRevLett.79.2604,PhysRevLett.81.933,santos2002collapse}.
The phenomenon, sometimes nicknamed ``Bosenova'' because of the subsequent explosion,
 has been demonstrated in a series of experiments which have attained a 
notable degree of control 
\cite{gerton2000direct,roberts2001controlled,donley2001dynamics,eigen2016observation}
and modeled using GPE
\cite{saito2001intermittent,saito2001power,adhikari2002mean,savage2003bose,ueda2003consistent}.

There is a vast literature on collapses in GPE/NLSE,
comprising analytic results, numerics and experiments
\cite{berge1998wave,sulem2007nonlinear,fibich2015nonlinear}.
Altogether, the standard lore is that collapse happens as soon as some 
parameter of the initial state,
for example the overall energy or the  peak value,   surpasses some threshold.
Near collapse, solutions typically display self-similarities \cite{budd1999new,sulem2007nonlinear}.

In a totally different arena, collapses have also been studied in general relativity (GR),
in relation to black hole (BH) formation. In asymptotically flat
space,  a critical value for a given parameter 
marks the limit between collapse and dispersion to infinity of the initial matter distribution \cite{christodoulou1986problem}. Remarkably, solutions at threshold
also display extremely interesting self-similarities both of  discrete and continuous type 
\cite{PhysRevLett.70.9,gundlach2003critical}.

In recent times, a modified version of this gravitational phenomenology 
has been explored in Anti de Sitter (AdS), the maximally symmetric spacetime with negative curvature.
The most dramatic difference is that
space becomes effectively bounded. Initial conditions
that do not form a BH right away, reach the boundary and bounce, 
falling back again towards the origin with a different radial profile. 
The surprising result of \cite{Bizon:2011gg} is that, no matter how 
small the initial amplitude, 
collapse eventually happens  after a  number of bounces of the matter field against the 
boundary.
The initial claim was that the underlying mechanism 
is an energy cascade triggered by a fully resonant spectrum of the linearized perturbations. Later, the  picture has proved to be more intricate and 
 has thereby become a subject of intense  research. The interested reader may consult 
 \cite{Maliborski:2013via,Deppe:2016gur} for further details.

It seems natural to look for 
similarities with
other nonlinear systems. In this work, we have
set up an efficient numerical code for radially symmetric long time simulations of the GPE
with attractive nonlinearity.  
In order to reproduce generic features of the scalar field  in  
 AdS, we consider a  quadratic potential, whose spectrum is fully resonant. Moreover, this potential
typically models BEC traps  and thus the framework is closely related
to Bosenovae.  Mathematical properties of the GPE in a harmonic trap has been studied elsewhere
\cite{carles2002remarks,jao2016energy,hani2015asymptotic}.
Our setup is similar in spirit to the evolution of a probe scalar in AdS \cite{Basu:2014sia}.

We perform series simulations in different number of dimensions $d$,
fixing the initial condition for the wavefunction to be a Gaussian  and studying the
outcome for varying amplitudes and widths. 
For large amplitudes, the attractive interaction is strong and causes a prompt collapse. 
Below  threshold for direct collapse, a 
singularity is reached after a number of oscillations. We refer to these as
``delayed collapses'' following the
nomenclature from the GR context \cite{Okawa:2013jba}.
We  see a stepwise structure of delayed collapses. 
In the physical $d=3$ case, our plots resemble those of wide initial gaussians in AdS \cite{buchel2013boson}, 
or those of theories with a
 mass gap for BH formation, such as AdS$_3$ \cite{daSilva:2016nah},
Einstein-Maxwell-scalar \cite{arias2016stability} or 
Einstein-Gauss-Bonnet \cite{deppe2015stability,Deppe:2016dcr}. There are plateaux
for which collapses occur after a number of bounces and transition 
regions where the collapsing time becomes a chaotic function of the initial amplitude.  
Below some amplitude, we do not find collapse in
our simulations. 
In $d=7$, our results become surprisingly similar to those of \cite{Bizon:2011gg}.

We close this introduction with a clarification. Our work is somewhat related, in spirit,
to analogue gravity \cite{barcelo2005analogue}, 
looking for systems that reproduce general relativistic phenomena
and discussing laboratory experiments that can mimic aspects of gravity, as {\it e.g.}
 \cite{Philbin1367} (see \cite{Hossenfelder1,Hossenfelder2} for a discussion
of Anti de Sitter in this context).
Nevertheless, we underscore that our approach is different.
We focus on the underlying nonlinear dynamics
of certain processes rather than looking for direct connections with the gravitational formalism
or the dynamics of probes in curved spaces. 
Despite analyzing very different equations, we will show that there are compelling
similarities in the results. Our work suggests that the remarkable  delayed collapse 
phenomenology recently found in
gravity might be more general than initially expected, opening the possibility of 
cross-fertilization between different fields of physics.

\section{II. Formalism}
 
The wavefunction $\psi(t,r)$, evolving  in a $d$-dimensional harmonic isotropic trap
$V(r)=\frac{r^2}{2}$ with cubic focusing nonlinearity, is governed by
\begin{equation}
i \partial_t \psi = -\frac12 \partial^2_r \psi- \frac{d-1}{2r}\partial_r \psi  + \frac12 r^2 \psi - |\psi|^2 \psi\, .
\label{GP}
\end{equation}
Without loss of generality, we use dimensionless quantities. In the
appendix, we give the well-known rescaling
from the dimensionful BEC variables for $d=3$.
The norm and hamiltonian are conserved
\begin{eqnarray}
N&=&S_d \int_0^\infty r^{d-1} |\psi|^2 dr \, ,
\label{norm}
\\
H&=&\frac{S_d}{2} \int_0^\infty r^{d-1} \left(
\partial_r \psi^* \partial_r \psi + \,  r^2 |\psi|^2 - |\psi|^4
\right) dr\, ,
\label{hamiltonian}
\end{eqnarray}
where $S_d$ is the  integral over the angles, {\it e.g.} $S_2=2\pi$, $S_3=4\pi$.
An important indicator  is the  variance 
$V=\langle r^2 \rangle_\psi=S_d \int_0^\infty r^{d+1}|\psi|^2 dr$. 
The value $V=0$  is tantamount to reaching collapse, in  that all the energy gets concentrated at the
origin \cite{sulem2007nonlinear}. However
 collapses can also occur with $V\neq 0$. 
A simple calculation shows that $V$ satisfies the
following virial identity, which is a straightforward generalization
of the standard one \cite{sulem2007nonlinear}:
\begin{equation}
\frac{d^2V}{dt^2}=4H-4 V - (d-2) S_d \int_0^\infty r^{d-1} |\psi|^4dr \, .
\label{variance}
\end{equation}
For our purposes, the significance of the variance is two-fold. First,
it allows to estimate how concentrated the wavefunction around the origin is and,
as stated above, it can be an indicator of collapse. Second, we can use the
identity (\ref{variance}) as a quality check for monitoring the accuracy of the numerical evolution (see the
appendix).

In order to study the  problem from the point of view of  the formalism of
weak turbulence \cite{PhysRevLett.106.115303,shukla2013turbulence},
 the evolution equation can be rewritten in terms of modes
 \begin{equation}
\psi(t,r)= \sum_{n=0}^\infty \alpha_n(t) e^{-i\,\mu^{(d)}_n t} f^{(d)}_n(r)\, ,
\label{modes}
\end{equation}
where the $f^{(d)}_n(r)$ are the
orthonormal
basis of eigenfunctions
of the linear problem which we give for completeness in the appendix.
Inserting (\ref{modes}) in (\ref{GP}), projecting on the $f_n^{(d)}(r)$ and
assuming that the rate of change of the $\alpha_n$ is much lower than that of the
complex exponentials, one finds
\begin{equation}
\dot \alpha_l(t) =  i \sum_{i=0}^\infty
\sum_{j=0}^\infty
\sum_{k=0}^\infty C_{ijkl} \alpha_i(t) \alpha_j(t) \alpha_k^*(t) \, ,
\label{modevol}
\end{equation}
where
\begin{equation}
C_{ijkl} = \delta_{i+j,k+l} S_d \int r^{d-1} f^{(d)}_i(r)f^{(d)}_j(r)f^{(d)}_k(r)f^{(d)}_l(r)dr\, .
\label{coefs}
\end{equation}
This multiscale analysis is standard  and goes under 
different names in the literature:
 rotating wave approximation,  two-time formalism \cite{Balasubramanian:2014cja}, 
 averaging \cite{Craps:2014vaa}, etc.  We find
The two conserved quantities (\ref{norm}), (\ref{hamiltonian}) 
(as in \cite{Basu:2014sia})
and the coincidence of the only resonant channel, $i+j=k+l$,
 relate closely the GPE formalism to the AdS setup \cite{Craps:2014jwa,Buchel:2014xwa}, and
imply the
coexistence of direct and inverse energy cascades. 

An important ingredient is the
 family of  initial conditions. We  use gaussians of  width $\sigma$ and
 amplitude $\epsilon$
\begin{equation}
\psi(t=0,r)=\epsilon \,e^{-r^2/\sigma^2}\, .
\label{ini}
\end{equation}
The  choice is motivated by simplicity and
the discussion is not significantly affected by this particular shape.
Eq. (\ref{ini}) gives the ground state of the linear problem in which the harmonic potential has an
 extra $4\sigma^{-4}$ factor. Hence one can think of the processes we simulate as quenches in which at $t=0$ the linear 
 and/or nonlinear potentials are abruptly modified initiating the dynamical evolution.
 In BECs, this is accomplished by tuning external fields which not only constitute the harmonic
 trap but do severely affect the atom-atom scattering length near Feshbach resonances 
 \cite{chin2010feshbach}.

\section{III. Numerical analysis}
 
Our work relies on  numerical integration of Eq. (\ref{GP}). The reader interested in the methods can find all the details in the appendix.  
We start by briefly commenting on $d=2$. In BECs, this limit is achieved with a strongly anisotropic
trap leading to a disk-shaped condensate \cite{gorlitz2001realization}.
In the absence of external potential, the Townes profile 
\cite{PhysRevLett.13.479,PhysRevLett.90.203902} marks the limit between directly collapsing
waves and those dispersing to infinity. The harmonic potential changes the picture, permitting stable
stationary solutions \cite{herring2008radially}. We have not found any set of initial
conditions yielding  the sought structure of delayed collapses and below threshold solutions remain 
regular for all times. 

\begin{figure}[h!]
\begin{center}
\includegraphics[width=\columnwidth]{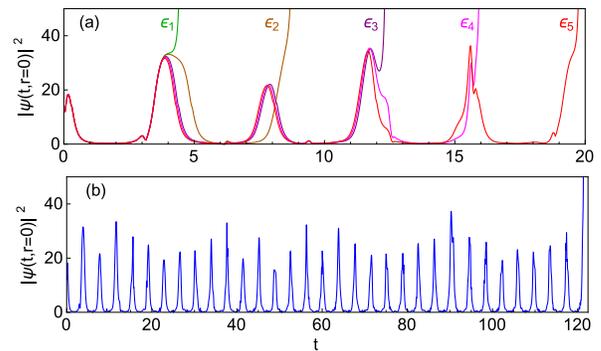}
\end{center}
\caption{Examples of delayed collapses in $d=3$. 
We depict the value of $|\psi(t,r=0)|^2$ 
for $\sigma^{2}=2/5$ and varying
$\epsilon$. In each case, the wavefunction promptly collapses after hitting the 
upper limit of the plot.
 In panel (a), ordered by growing $t_c$, we have $\epsilon_1=3.1286$
 (green line), $\epsilon_2$=3.1285  (brown), $\epsilon_3$=3.1280
  (purple), $\epsilon_4$=3.1277  (magenta) and $\epsilon_5$=3.1273  (red).
Notice the approximate overlap for long ranges of $t$.
In panel (b), $\epsilon_6=3.1270$ (blue line) results in collapse after many bounces.
We remark that the time scale of both graphs is different.
}
\label{fig1}
\end{figure}

In $d=3$ without trapping potential, there is also a sharp separation between direct collapse
and dispersion. 
With a harmonic potential, apart from having periodic stationary solutions 
\cite{PhysRevA.51.4704,PhysRevE.92.013201}, 
 delayed collapses are possible. 
Figure \ref{fig1} shows several examples.
Each line of the plot is found by numerically integrating Eq. (\ref{GP}) with
initial conditions (\ref{ini}). The different initial conditions have all the same 
width $\sigma$ but different
values of the height $\epsilon$. We represent the value of $|\psi|^2$ at the origin.
Collapse happens when this quantity diverges. For each initial condition, we can
therefore compute
the time of collapse $t_c$. There is an oscillating behavior which can be heuristically
understood in terms of the linear problem (see the appendix), for which it can be immediately checked
 that $|\psi(r)|^2$ returns to itself with time period $\pi$. The same period is directly
read from Eq. (\ref{variance}), since the last term is not present in the linear case.
When the nonlinear term is present, the oscillation is only approximate
and the ``period'' for which maxima appear in Fig. \ref{fig1} is not exactly $\pi$.
At each bounce inside the harmonic potential, the spatial profile of $|\psi|^2$ is
modified and, eventually, this can result in a delayed collapse after several or many swings. 
This number of oscillations can
be adjusted by fine-tuning the initial conditions.
We call delayed collapses those which happen after one or more oscillations,
like all the examples in the figure.

\begin{figure}[h!]
\begin{center}
\includegraphics[width=\columnwidth]{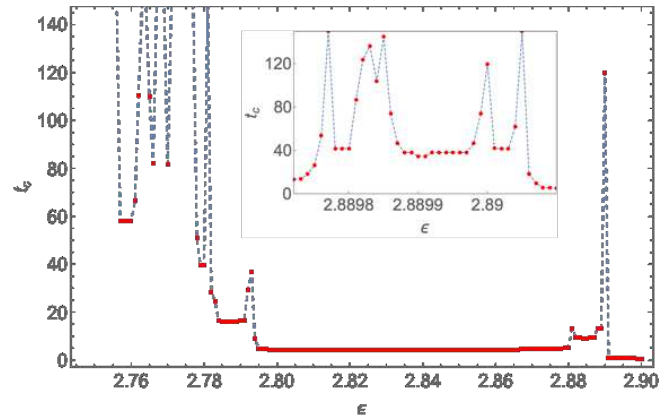}
\end{center}
\caption{Time of collapse for a family of initial conditions  with $\sigma^2 = 1/2$
in $d=3$. Prompt collapse occurs for $\epsilon>2.891$. The resolution of the peak
at the transition  $\epsilon\sim 2.889$ exhibits a chaotic structure (in the inset).}
\label{fig2}
\end{figure}

In figure \ref{fig2}, we plot the time of collapse, $t_c$, as a function of $\epsilon$ for 
fixed $\sigma^{2}=1/2$. 
Each red dot corresponds to a full simulation of Eq. (\ref{GP}) for which we find
$t_c$ as in Fig. \ref{fig1}. 
Necessarily, the computation has to be stopped at some value of $t=t_{comp}$.
If collapse has not been found by that time, it means that $t_c>t_{comp}$ or, possibly,
there is no collapse in any finite time. In Fig. \ref{fig2} we have taken $t_{comp}=140$ and
the non-collapsing points are represented by the dotted line going up the top of the graph.

There are three distinct regions in parameter space. 
For large $\epsilon$ ($\epsilon>2.891$ in the case of Fig. \ref{fig2}), there is direct collapse and
$t_c$ decreases with growing $\epsilon$.
For  small $\epsilon$ ($\epsilon < 2.755$ in the case of Fig. 2),
solutions stay regular for $t<t_{comp}$. There is a 
noteworthy  intermediate
transition region where  some step structure is apparent (cf. 
$2.755<\epsilon<2.891$ in Fig. \ref{fig2}). Roughly
speaking, each step corresponds to a number of bounces in the harmonic potential. 
At the boundary between steps, $t_c$ presents a bump. 
Our plot  shows remarkable features in common with those found in different AdS setups (cf. Figs. 9 
in \cite{buchel2013boson},  3 in \cite{cardoso2016collapsing}, 16 in \cite{arias2016stability} or 
2 in \cite{deppe2015stability}). The chaotic character of the curve 
at the bumps has been recently  established in \cite{Deppe:2016dcr}.
An analogous
 detailed analysis for this case would be of interest but extends way beyond the scope of this paper and is left for future work.
The common feature of these examples seems to be the lack of fully resonant
 linearized perturbations around a standing wave \cite{Maliborski:2014rma}.

\begin{figure}[h!]
\begin{center}
\includegraphics[width=\columnwidth]{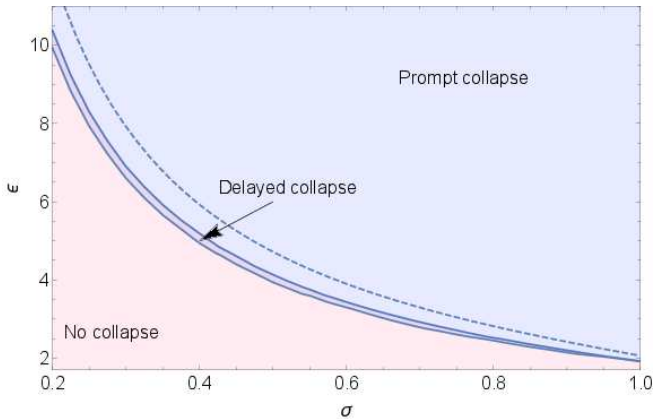}
\end{center}
\caption{Sectors for prompt and delayed collapse in $d=3$. 
The dashed line signals the transition from 
initial conditions where $d^2V/dt^2|_{t=0}$ changes from negative (above) to positive (below). All delayed collapses happen 
with $V\neq 0$.}
\label{fig3}
\end{figure}

Figure \ref{fig3} depicts a map of the results obtained with different $\epsilon$ and $\sigma$,
found by repeated simulations of Eq. (\ref{GP}) with different initial conditions 
(notice that the computations leading to Fig. \ref{fig2} correspond to a vertical line
at fixed $\sigma=0.707$ around the delayed collapse region).
The transition line between prompt and delayed collapse can be defined with precision. 
On the other hand, regularity of evolution can only be stated within a given computational time, which
we have fixed to $t_{comp}=100$ for Fig. \ref{fig3}. 
The delayed collapse window is certainly narrow in the two-parameter space of initial conditions. 

Tuning  $\epsilon$ is equivalent to tuning the scattering length in the cold atom
framework. Thus, Fig. \ref{fig2} depicts
the same observable as, {\it e.g.}, Fig. 2 of \cite{donley2001dynamics}.
Our work is devoted to the theoretical analysis of Eq. (\ref{GP}) and we
do not intend here to make a realistic description of experimental situations, where, typically, 
anisotropies and nonlinear losses should be taken into account. In any
case, it is interesting 
to make order of magnitude estimates of the parameters. We take 
Fig. 2 of \cite{donley2001dynamics} for reference. There, the harmonic trap
is mildly anisotropic with $\nu_{radial}=17.5$Hz and $\nu_{axial}$=6.8Hz. Let us
take a fiducial value of that order of magnitude $\nu=14$Hz. The initial condition 
is the ground state of the linear problem, namely $\sigma = \sqrt2$. In Fig. \ref{fig3},
it is shown that the delayed collapse window in parameter space closes around $\sigma=1$
and therefore the original atom cloud of \cite{donley2001dynamics} is too wide to
produce the behavior described here. Let us however compute the range of $\epsilon$
corresponding to the horizontal axis of Fig. 2 of \cite{donley2001dynamics}.
The relation between the physical and the adimensional quantities is provided in the appendix
and we take $\bar N=6000$ for the number of atoms and $m=1.4\times 10^{-25}$kg (for $^{85}$Rb).
It is straightforward to show that $4 < |a|/a_0< 60$, where $a_0$ is the Bohr radius corresponds
to $1 < \epsilon<3.8$. 
The situation described in the present paper could be thus approached
by taking an isotropic trap and starting with a narrower profile. Experimentally, smaller values
of the initial $\sigma$ should be attainable by taking a tighter trap for $t<0$.
At $t=0$, the trap can be ramped down while the value of $|a|$ is ramped up.
Imagine a situation with $\sigma^2=1/2$ and the
rest of parameters fixed as above. The $2.75<\epsilon<2.90$ window of our Fig. \ref{fig2} 
becomes $3.8 < |a|/a_0< 4.3$ in physical terms, values that
are well within reach with Feshbach resonance techniques.

As mentioned above, our interest in looking for delayed collapses
in GPE was sparked by the similar 
problem in the gravitational context 
of a scalar field coupled to gravity and a negative cosmological constant. Now we can reverse the lore, and 
ask ourselves what we can  learn
from this  nonlinear system in regard to the important question of the instability of AdS. In fact, one 
 advantage of this simple equation 
is the fact that we can tweak separate features independently, like the  property of full resonance or the character of
the nonlinearity. These features are
utterly mixed in the gravitational setup and are difficult to disentangle \cite{Cai:2015jbs}.

There is growing consensus  that the weak turbulent instability of AdS, 
{\it i.e.}, that  initial data always collapse in the limit $\epsilon \to 0$,
is not solely caused by a fully  resonant spectrum. This being a necessary condition,  needs to be  supplemented with appropriate asymptotics for the
coupling coefficients $C_{ijkl}$  (\ref{coefs}) at  large  values of $i,j,k,l\to \infty$ in a resonant channel. 
This should catalyse  efficient energy transfer to the  high frequency  modes.
In the three-dimensional GPE,  we have checked that  $C_{in,jn,kn,ln}$ behaves with  $n\to \infty$ as a power $n^\gamma$ with $\gamma = -0.5$.
This is substantially lower a growth than the one observed in $AdS_{4}$ where $\gamma = 1$ 
in this channel  \cite{Craps:2015iia} (in order to compare the resonant system in 
AdS with ours in Eq. (\ref{modevol}), the modes $\alpha_i$ need to be rescaled to $\alpha_i/\sqrt{\omega_i}$. This shifts $\gamma=d \to d-2$). 

 A natural question is what  minimal twist  could we perform   in order
 to enhance the asymptotics, and whether this would have the expected impact on the collapse at low values of the initial amplitude. 
   One possibility is to include derivative terms, (derivative couplings appear in the effective equations for a scalar field coupled
to gravity). In fact, a  term of the form $|\partial_r \psi|^2 \psi$ added to the GPE  yields  $\gamma \approx 0.5$. The numerical evolution of the system with 
  this  additional term  becomes  unstable and we have deferred its study. 
  
\begin{figure}[h!]
\begin{center}
\includegraphics[width=\columnwidth]{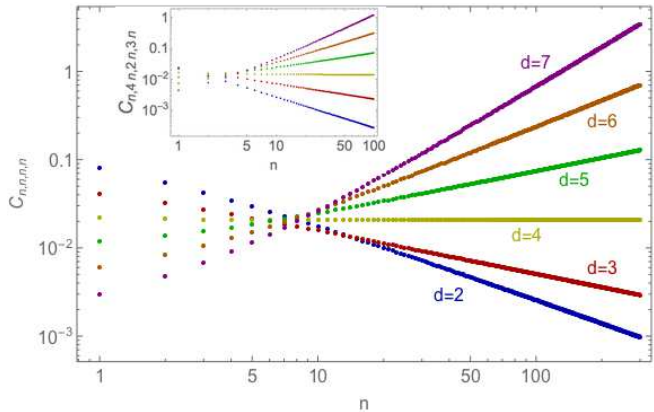}
\end{center}
\caption{Doubly logarithmic plot of $C_{nnnn}$ as a function of $n$ 
for different values of $d$. In the inset, we depict the $C_{n,4n,2n,3n}$, showing that
they present a
similar behavior.}
\label{fig4}
\end{figure}

\begin{figure}[h!]
\begin{center}
\includegraphics[width=1.\columnwidth]{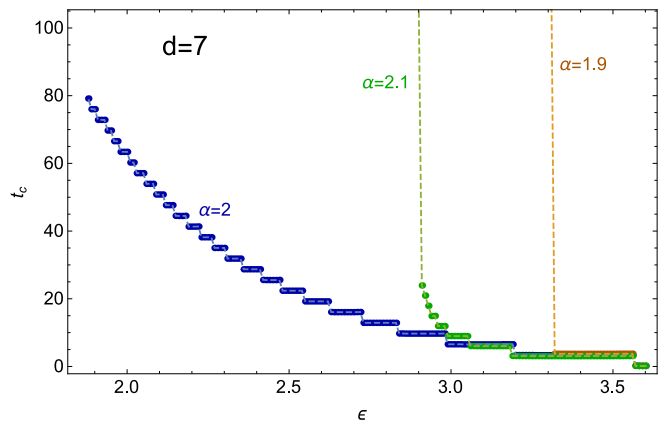}
\end{center}
\caption{Curve of $t_c$ vs. $\epsilon$ for $d=7$ with $\sigma^2=\frac12$ and 
$V(r)=\frac12 r^\alpha$. 
Two non-resonant cases $\alpha=1.9$ and $\alpha=2.1$ present limiting values
of $\epsilon$ below which no singularity is reached in computational time.
The harmonic resonant case $\alpha=2$ (Eq. (\ref{GP})) shows a regularly 
spaced set of steps extending to much lower
values of $\epsilon$.
 }
\label{fig5}
\end{figure}

  Another artificial but efficient way to enhance $\gamma$
 is to formally extend the GPE to higher dimensions.  In figure \ref{fig4}, 
 we depict the behavior of 
 $C_{nnnn}$ for
 $n \leq 300$ and $d=2,\dots,7$. In the large $n$ limit  the exponent  $\gamma$ behaves with the dimension as
 (see the appendix)
 \begin{equation}
 \gamma  = \frac{d}{2}-2,
 \label{gama}
 \end{equation}
 a fact  that presumably causes the different qualitative behavior between  $d=2$ and $d=3$. 
 This suggests that 
 turbulence
and collapse are favored at large $d$.   
This expectation is borne out by the results of figure \ref{fig5}, which shows
a neat step structure for $t_c(\epsilon)$ when $d=7$, resembling  $AdS_4$  \cite{Bizon:2011gg}.

We also display the results with a slightly modified potential $V(r)=\frac12 r^\alpha$, where
$\alpha=2$ corresponds to the harmonic case. Changing $\alpha$ modifies the eigenvalues of the linear
problem, breaking  full resonance. As shown in fig. \ref{fig5} this has a dramatic
 impact on the time of collapse
at low initial amplitudes.

\section{IV. Summary of results and connections to AdS gravity}

Collapses in GPE/NLSE  have been thoroughly studied but  are still a fascinating topic. 
We have considered  a harmonic trap in different number of dimensions. 
We have studied the long time evolution of the wave function
for families of gaussian initial conditions depending on two parameters.
For long time evolution we mean that many bounces are allowed to take place
 by evolving up to a time $t_{comp} \gg \pi$, which
is much larger than that of an oscillation in the potential (we have seen that nonlinear
oscillations do not have exactly a period $\pi$ but are associated
to time spans of that order).
 In $d=3$, we have seen that there
are initial conditions that lead to delayed collapses, where the divergence occurs after a number of 
oscillations, Fig. \ref{fig1}. 
On  average, the time of collapse grows with the inverse of the 
norm of the initial wavefunction, but non-monotonically and with a strong sensitivity to the initial condition.
This happens in a region of parameter  space which separates direct
collapse from absence of collapse, Figs. \ref{fig2} and \ref{fig3}. 
In $d=7$, delayed collapses are also present, albeit of a different type.
A neat step structure appears, cf. Fig. \ref{fig5}, which extends collapse to smaller
values of $\epsilon$. This happens thanks to a fully resonant spectrum and 
efficient energy transfer between modes, associated to rapidly growing $C_{ijlm}$ couplings,
Fig. \ref{fig4}.

\begin{figure}[h!]
\begin{center}
\includegraphics[width=1.\columnwidth]{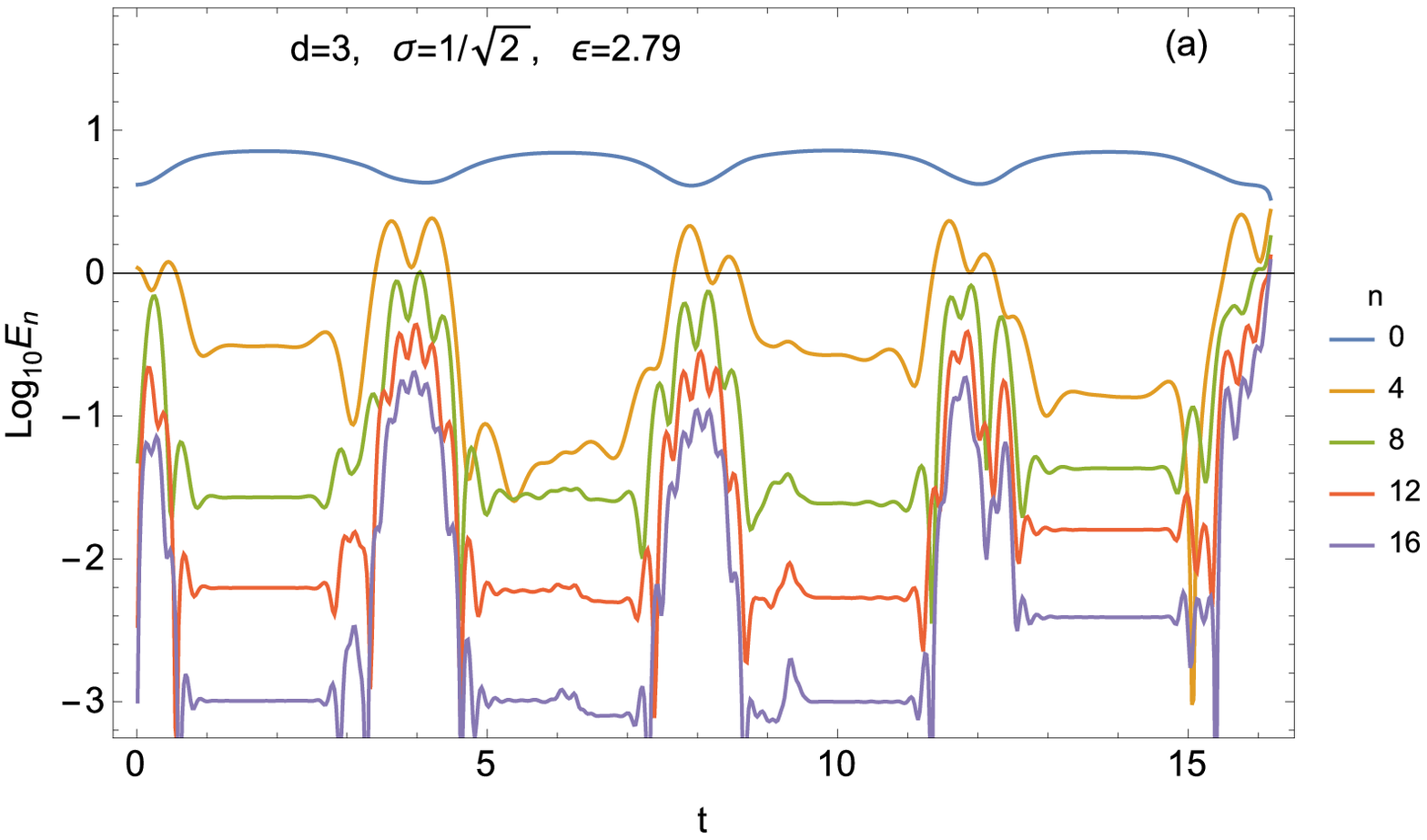}
\includegraphics[width=1.\columnwidth]{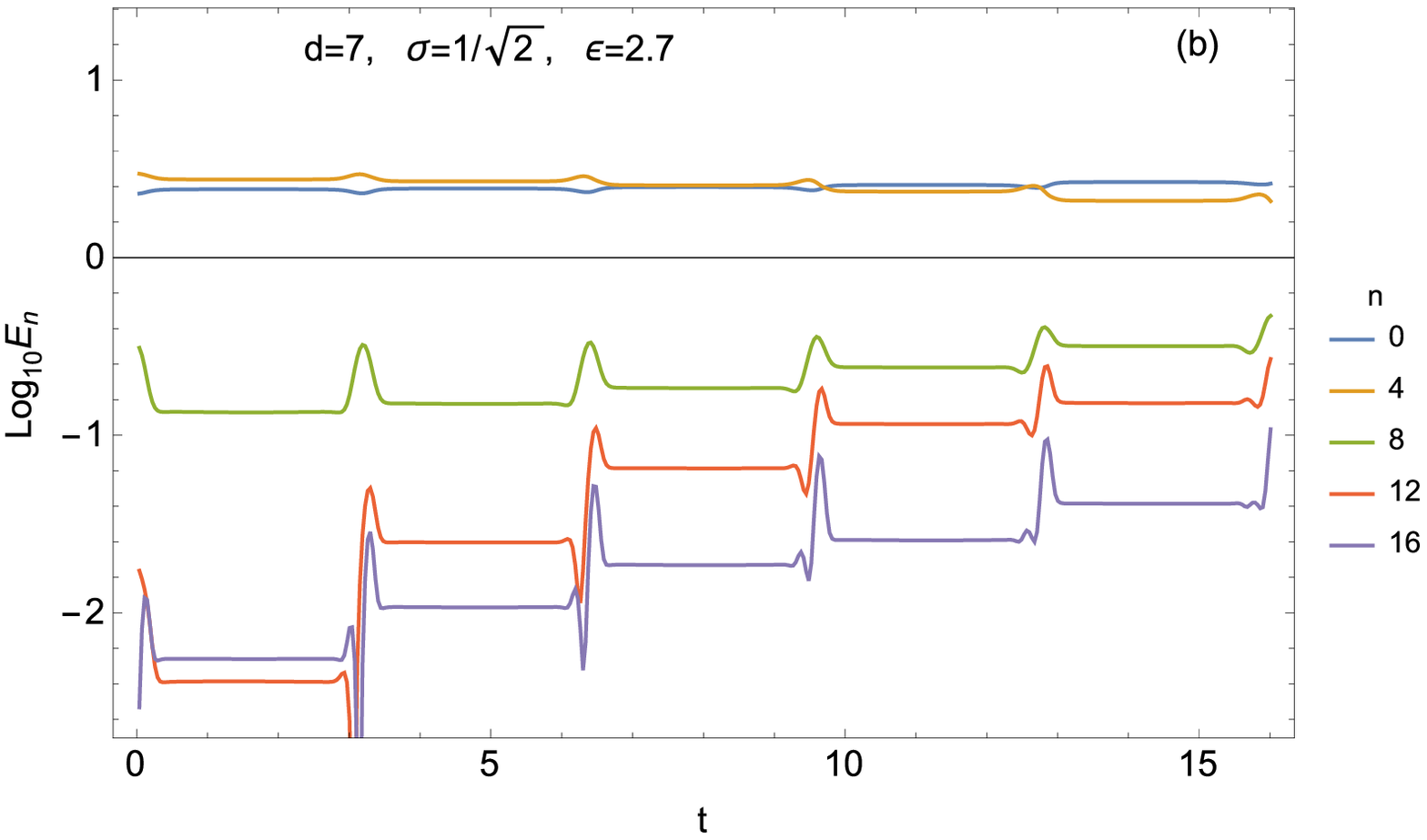}
\end{center}
\caption{Temporal evolution of the spectrum $E_n=\mu_n |\alpha_n|^2$ for 
some low-lying modes until 
collapse.
Panel (a) corresponds to an example with $d=3$.
Panel (b) corresponds to an example with $d=7$.
The changes in the spectrum mostly take place 
around the time when the wavefunction periodically peaks at the
origin.  This periodic energy transfer towards the ultraviolet underlies the stepwise
patterns of collapse in figures \ref{fig2} and \ref{fig5}. 
 }
\label{fig6}
\end{figure}

For completeness, in Fig.\ref{fig6} 
we have plotted the behavior of the energy spectrum as a function of time for some of the
 modes. We do not include more modes for clarity of the plot.
Besides the usual direct cascade
that feeds the higher modes from the lowest ones,  we also observe an inverse cascade.
This is a consequence of the simultaneous conservation of two
independent quantities, $N$ and $H$, and matches the similar behaviour observed in the resonant system 
obtained from averaging the AdS dynamics (see section III in \cite{Buchel:2014xwa}).

The transitions between steps in Fig. \ref{fig6} happens at the time when the pulse comes close to the origin. A similar structure is seen in AdS and stems from the fact that the magnitude of the wave function increases by orders of magnitude, enhancing the non-linear effects at those instants of time. In $d=7$, this process is very much ordered and seems to end up inexorably in a collapse whereby a polinomially decaying spectrum is achieved. In $d=3$ the ordered structure is less apparent and the process becomes a statistical wave turbulence effect. 

One may wonder is how generic the described behavior is.
One intriguing answer is given by comparing to a scalar field in AdS gravity which shows
remarkable similarities. 
Indeed, the two-fold structure with (partially) chaotic and 
ordered delayed collapses
has been described in that context, see the references
provided in the previous section. 

In summary,  we have found an interesting structure in the space of evolutions leading to collapse, within
the  simple setup of Eq. (\ref{GP}). There are neat similarities 
with those encountered in the context of gravitational dynamics in AdS. Since the inherent physics and the
 evolution equations are 
rather different in both cases, a natural question is what underlying mechanisms
are at work. Both  cases have an attractive nonlinearity that can bunch together 
the energy and produce a collapse that alters the otherwise smooth evolution.
The harmonic potential and the AdS boundary act as boxes that impede the
escape of the perturbation to infinity and cause the bounces. 
Moreover, they both provide a fully resonant
spectrum for the linear modes. In both cases, 
there is an effective equation (\ref{modevol})
and the resonant channels and conserved quantities coincide.
These observations might be useful to identify other frameworks in which
similar phenomena  take place. Nevertheless, solid conclusions can only
be derived from full-fledged
numerical analysis.

Turbulence and energy cascades constitute an active
area of research in the context of GPE, see {\it e.g.} \cite{shukla2013turbulence}
\cite{carles2010energy,turbu1, turbu2}, and especially \cite{Tsatsos2016} for a recent review with references.
Our setup differs from these works in several aspects, including the presence of the harmonic term
and the assumed radially symmetric evolution.
It could be interesting to explore
whether insights from the quoted studies have implications in the present context.

\section{V. Outlook}

The GPE/NLSE is an ubiquitous formalism in nonlinear waves,
and this fact can pave the way for tabletop experiments with
qualitative
 analogies to AdS gravity.
The  experimental control reached in  BECs might provide an ideal
framework to
 realize the delayed collapse phenomenology.
 Our formalism is not a complete description of any such system since, for instance,
it does not include
anisotropies or losses by three-body recombinations.
However, our results suggest that it might be worth revisiting, theoretically and/or experimentally,
the transition from a stable BEC with attractive nonlinearity to a 
Bosenova collapse (cf. Fig. 2 of \cite{donley2001dynamics}).
Depending on the width of the initial wavefunction, $t_c$ can decrease with
the strength of the interaction
$|a|$, a counterintuitive behavior. Moreover, there are narrow regions of parameter space where
$t_c$ changes chaotically. Whether these properties stemming from Eq. (\ref{GP})
can be consequential in a realistic setup is a challenging question for the future.

Our model can also be used as a mathematical tool to
provide new insights to the problem of  AdS instability.
The  GPE evolution equation is much simpler than those of AdS gravity 
but, as we have discussed,  
its study might be 
instrumental in understanding
which peculiarities of general relativity are essential for the delayed collapse behavior and which are
shared by other kinds of nonlinear systems. 
Moreover, it provides turning knobs  that can be used
to investigate which qualitative features ({\it e.g} full resonance, efficient turbulent cascades, etc.)
underlay the remarkable nonlinear dynamics that has been discovered in the AdS context in
recent years. For instance, the  behavior of the
$C_{ijkl}$ coefficients (\ref{coefs}) can be tuned by considering larger dimensions, nontrivial
spatial profiles for the nonlinear coupling or other generalizations of Eq. (\ref{GP}).
On the other hand, full resonance can be broken by slightly modifying the potential. 
We hope that a detailed scrutiny of these issues will shed new light on this fascinating 
nonlinear dynamics.

\appendix

\section*{Appendix}

In this appendix we lay out some technicalities related to the main text.
We write down some useful well-known 
equations concerning the dimensionful
Gross-Pitaevskii equation (GPE) and the radially symmetric
eigenmodes of a harmonic potential in $d\geq 2$ dimensions. 
Then, we specify some details regarding the employed numerical methods and
cross-checks. Finally, we explain how the asymptotic form of the $C_{ijkl}$ 
coefficients can be derived from the integrals.

\subsection{ The $d=3$ dimensionful GPE}

The Gross-Pitaevskii mean-field description of a dilute Bose-Einstein
condensate of bosons of mass $m$ in an isotropic harmonic potential of
frequency $\omega$ is:
\begin{equation}
i \hbar \partial_{\tilde t} \tilde \psi = - \frac{\hbar^2}{2m}\tilde \nabla^2 
\tilde \psi + \frac{m\omega^2}{2}\tilde r^2 \tilde \psi + \frac{4\pi \hbar^2a}{m} |\tilde \psi|^2 \tilde \psi
\label{GPdim}
\end{equation}
where $\tilde \nabla^2$ is the $3-$dimensional Laplacian, $a$
is the s-wave scattering length which can be positive (repulsive interaction)
or negative (attractive interaction) and $\tilde N= \int |\tilde \psi|^2 d^3 \tilde {\bf r}$ corresponds to
 the number of bosons in the sample.  
Eq. (\ref{GP}) is recovered by appropriately rescaling $\tilde t$, $\tilde r$ and
$\tilde \psi$. The relation between normalizations is
$\tilde N = \frac{1}{4\pi |a|} \sqrt\frac{\hbar}{m\omega}N$.

\subsection{Eigenfunctions of the harmonic oscillator}

The angle-independent  
eigenfunctions of the linear quantum harmonic oscillator 
with $d\geq 2$
are
\begin{equation}
\psi_n= e^{-i\mu_n^{(d)}t}f_n^{(d)}(r)\, 
\end{equation}
with a fully resonant spectrum $\mu^{(d)}_n=2n + d/2$ and
\begin{equation}
f_n^{(d)}(r) =  \sqrt\frac{n!\Gamma(\frac{d}{2})}{\pi^{d/2}\Gamma(\frac{d}{2}+n)}
 \  L_n^{(d-2)/2}(r^2) e^{-r^2/2}\, .
 \label{fndr}
\end{equation}
The $L_n^{(d-2)/2}$ are generalized Laguerre polynomials,
$\Gamma$ represents Euler's gamma function and the
multiplicative  constant is chosen to satisfy the orthonormality condition $\int  f_n^{(d)}f_{m}^{(d)}d^d{\bf r} = \delta_{nm}.$

\subsection{ Numerical details and quality checks}

In the region of delayed collapses, small changes in initial conditions can lead to 
rather different results. Thus, a method for fast, stable and precise computation is needed and
we briefly describe here the one we have used.
Our integration algorithm relies on an explicit finite difference scheme consistent with fourth order accuracy
and convergence.
Spatial derivatives are discretized with standard stencils, and time evolution uses a
fourth order Runge-Kutta. The Courant factor, $c$,  depends on the discretization density since 
the equation is parabolic. Stability enforces this number to be quite small. For $2^w$ points
in the spatial grid,   we have used $c= 0.004\times 2^{12-w}$ with 
$w=12,...,17$. 

At the origin, we ensure regularity  by enforcing $\partial_r \psi (t,r=0)=0$. 
The harmonic potential confines the wavefunction and therefore $\psi$ decays 
exponentially with $r$. We have checked that choosing $r_{max}=50$ 
keeps $\psi(r_{max}) \sim 10^{-15}$ so that setting its value to zero, and
truncating $r$ to a finite interval $0<r<r_{max}$,
does not affect the simulation.  We have also performed computations using  a truly compact coordinate 
$r= z/(z_{max} - z)$ and found consistent results.

\begin{figure}[h!]
\begin{center}
\includegraphics[width=\columnwidth]{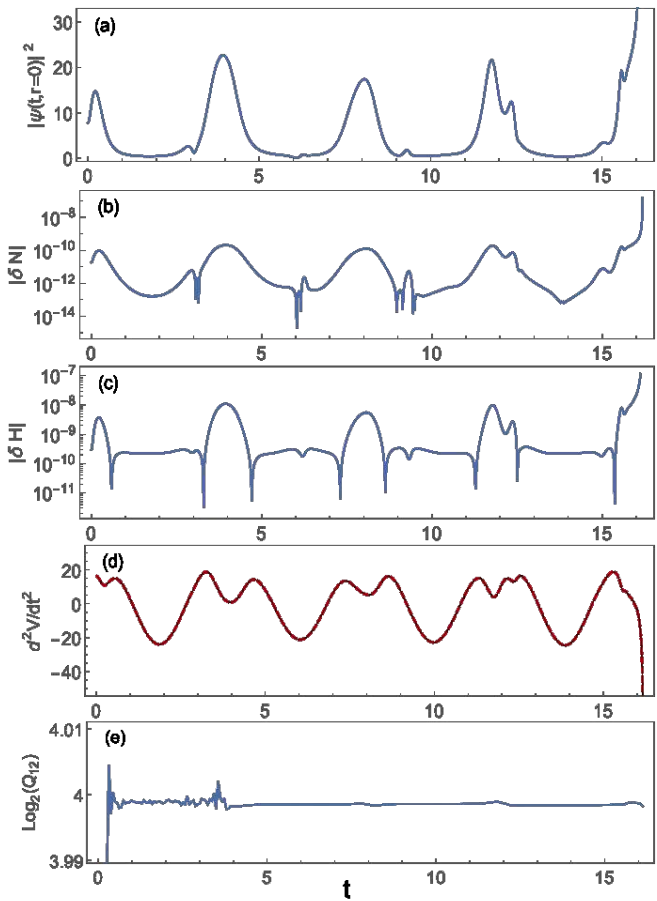}
\end{center}
\caption{A simulation with $d=3$, $\epsilon=2.79$  , $w=14$, $\sigma^2=1/2$.
Panel (a) represents  $|\psi|^2$ at the origin. Panels (b) and (c) depict the relative
deviations for the norm and hamiltonian, respectively. Panel (d) compares
the  second derivative of the variance from the numerical computation to that from the
right-hand side of Eq. (\ref{variance}) computed at each step.
Finally, panel (e) shows the convergence level along the whole simulation. On a grid of $2^w$ points, 
$Q_w = ||\psi_{w+1}-\psi_w ||/||\psi_{w+2}-\psi_{w+1} ||$ where $||\cdot ||$ stands for a $L^2$ norm. The plot shows convergence at fourth order with 
high accuracy.  The tests show that the numerical method is reliable until times very near collapse.
}
\label{fig7}
\end{figure}

We have used the conservation of the norm (\ref{norm}), the  hamiltonian (\ref{hamiltonian}),
and the variance identity (\ref{variance}), as quality tests to monitor the  computation. 
For initial conditions of the form (\ref{ini}), they take the
values
\begin{eqnarray}
N&=&\left(\frac{\pi}{2}\right)^\frac{d}{2} \epsilon^2 \sigma^d  \, ,\nonumber
\\
H&=&\frac{d}{2}\left(\frac{\pi}{2}\right)^\frac{d}{2} \epsilon^2 \sigma^d \left(\frac{1}{\sigma^2}+\frac{\sigma^2}{4}
-\frac{\epsilon^2}{ 2^\frac{d}{2}d}\right)\, .
\label{HN}
\end{eqnarray}
The value of the second derivative of the variance at $t=0$ is:
\begin{equation}
\left.\frac{d^2V}{dt^2}\right|_{t=0}=2d\left(\frac{\pi}{2}\right)^\frac{d}{2} \epsilon^2 \sigma^d \left(\frac{1}{\sigma^2}-\frac{\sigma^2}{4}
-\frac{\epsilon^2}{ 2^{\frac{d}{2}+1}}\right)\, .
\end{equation}
We define the relative deviations $\delta N= \frac{N_{num}-N}{N}$,
$\delta H= \frac{H_{num}-H}{H}$ where $N_{num}$, $H_{num}$ are the values
obtained from numerical integration and $N$, $H$ are the values given in 
(\ref{HN}).
 To keep them satisfied at relative orders $\delta H, \delta N <10^{-6}$ we
have  implemented  global (in space) refinement. This is seen to be required  when the profile becomes extremely sharp at the origin, and needs more resolution to keep numerical control.
Fig. \ref{fig7} depicts a check for one of the simulations yielding a delayed collapse.

We have implemented our codes for CPU  computation and, in order to speed up the simulation, we
 also adapted them for GPGPU. The computation times that we achieve are approximately one 
 order of magnitude lower
in the latter case.

\subsection{Asymptotic values of the $C_{ijkl}$}

We now obtain the asymptotically large $n$ behaviour of 
  $C_{n\left(ijkl\right)}$. The form of the coefficients is
  given by the integral in (\ref{coefs}) for the eigenfunctions of  Eq.
  (\ref{fndr}). We follow the procedure of \cite{Evnin:2016mjx}.
Consider the identity \cite{Digital}
\begin{equation}
L_{n}^{(\alpha)}(\nu x)\approx\frac{e^{\frac{1}{2}\nu x}}{2^{\alpha}x^{\frac{1}{2}\alpha+\frac{1}{4}}}\left(\xi^{1/2}J_{\alpha}(\nu \xi)\right) 
 \label{eq:laguerre_as_bessels_fun}
\end{equation}
valid in the limit $n \to \infty$, $x\ll 1$. 
$J_{\alpha}(x)$ are Bessel functions of the first kind and  $\alpha=\frac{d}{2}-1$, 
$\nu=2\mu_n^{(d)}$, $\xi\approx \sqrt{x}$.
Taking $r^{2}=\nu x$, we get an expression for the eigenfunctions. 
Notice that the limit $x\ll 1$ 
is justified because $x= r^{2}/n$ with $n\rightarrow \infty$. 

Now the  Bessel functions are $J_{\alpha}(\nu^{1/2}r)$ and performing a
 change of variables in the integral (\ref{coefs}), we can remove the dependence in $n$ from
  the argument. Using the asymptotic value for the normalization 
 constants (cf. the prefactors in (\ref{fndr})) $N_n \sim n^{-\alpha/2}$,
  we obtain $C_{n\left(ijkl\right)} \sim n^{\frac{d}{2}-2}\int dz K(z)$ where $K(z)$
   does not depend on $n$. 
  Thus, we get equation (\ref{gama}).

We have checked that this expression is in  agreement with the values 
 of $C_{nnnn}$ and $C_{n,4n,2n,3n}$ found by numerical integration
for $d=2,\dots,7$, see figure \ref{fig4}.

\

\acknowledgements
{\it Acknowledgements.} We thank P. Carracedo, B. Craps, O. Evnin,
A. Rostworowski and A. Serantes for
useful comments.
This work was supported by grants FPA2011-22594, FIS2014-58117-P and FIS2014-61984-EXP
from Ministerio de Ciencia e Innovaci\'on and grants GPC2015/019, GRC2013/024, EM2013/002 
and AGRUP2015/11 from Xunta de Galicia. Part of this research has benefited from the use computational resources/services provided by the Galician Supercomputing Centre (CESGA)


\begin{thebibliography}{10}

\bibitem{agrawal2007nonlinear}
G.~P. Agrawal, {\em Nonlinear fiber optics}.
\newblock Academic press (2007).

\bibitem{malomed2005spatiotemporal}
B.~A. Malomed, D.~Mihalache, F.~Wise, and L.~Torner, 
   J. Opt. B: Quantum Semiclass. Opt. {bf 7}, R53,
  (2005).

\bibitem{zakharov2012kolmogorov}
V.~E. Zakharov, V.~S. L'vov, and G.~Falkovich, 
{\em Kolmogorov spectra of
  turbulence I: Wave turbulence}.
\newblock Springer Science \& Business Media (2012).

\bibitem{schive2014cosmic}
H.-Y. Schive, T.~Chiueh, and T.~Broadhurst, 
Nat. Phys. {\bf 10}, 
  496 (2014).

\bibitem{paredes2016interference}
A.~Paredes and H.~Michinel, 
 Phys. Dark Univ. {\bf 12}, 50 (2016).

\bibitem{dalfovo1999theory}
F.~Dalfovo, S.~Giorgini, L.~P. Pitaevskii, and S.~Stringari, 
 Rev. Mod. Phys. {\bf 71}, 463 (1999).

\bibitem{carretero2008nonlinear}
R.~Carretero-Gonz{\'a}lez, D.~Frantzeskakis, and P.~Kevrekidis, 
    Nonlinearity {\bf 21}, R139 (2008).

\bibitem{sulem2007nonlinear}
C.~Sulem and P.-L. Sulem, {\em The nonlinear Schr{\"o}dinger equation:
  self-focusing and wave collapse}, vol.~139.
\newblock Springer Science \& Business Media (2007).

\bibitem{PhysRevLett.79.2604}
Y.~Kagan, E.~L. Surkov, and G.~V. Shlyapnikov, 
   Phys. Rev. Lett. {\bf 79}, 2604 (1997).

\bibitem{PhysRevLett.81.933}
Y.~Kagan, A.~E. Muryshev, and G.~V. Shlyapnikov, 
  Phys. Rev. Lett. {\bf 81}, 933 (1998).

\bibitem{santos2002collapse}
L.~Santos and G.V.~Shlyapnikov, 
  Phys. Rev. A {\bf 66}, 011602 (2002).

\bibitem{gerton2000direct}
J.~M. Gerton, D.~Strekalov, I.~Prodan, and R.~G. Hulet, 
  Nature, {\bf 408}, 692 (2000).

\bibitem{roberts2001controlled}
J.~L. Roberts, N.~R. Claussen, S.~L. Cornish, E.~A. Donley, E.~A. Cornell, and
  C.~E. Wieman, 
  Phys. Rev. Lett. {\bf 86}, 4211 (2001).

\bibitem{donley2001dynamics}
E.~A. Donley, N.~R. Claussen, S.~L. Cornish, J.~L. Roberts, E.~A. Cornell, and
  C.~E. Wieman, 
 Nature {\bf 412}, 295 (2001).

\bibitem{eigen2016observation}
C.~Eigen, A.~L. Gaunt, A.~Suleymanzade, N.~Navon, Z.~Hadzibabic, and R.~P.
  Smith, 
Phys. Rev. X {\bf 6}, 041058 (2016).

\bibitem{saito2001intermittent}
H.~Saito and M.~Ueda, 
  Phys.
  Rev. Lett. {\bf 86}, 1406 (2001).

\bibitem{saito2001power}
H.~Saito and M.~Ueda, 
   Phys. Rev. A {\bf 63}, 043601 (2001).

\bibitem{adhikari2002mean}
S.~K. Adhikari, 
  Phys. Rev. A {\bf 66}, 013611 (2002).

\bibitem{savage2003bose}
C.M.~Savage, N.P.~Robins, and J.J.~Hope, 
   Phys. Rev. A {\bf 67},
  014304 (2003).

\bibitem{ueda2003consistent}
M.~Ueda and H.~Saito, 
   J. Phys. Soc. Jpn. {\bf 72}, 127
 (2003).

\bibitem{berge1998wave}
L.~Berg{\'e}, 
  Phys. Rep. {\bf 303}, 259 (1998).

\bibitem{fibich2015nonlinear}
G.~Fibich, {\em The nonlinear Schr{\"o}dinger equation}.
\newblock Springer (2015).


\bibitem{budd1999new}
C.~J. Budd, S.~Chen, and R.~D. Russell, 
  J.  Comput. Phys. {\bf 152}, 756 (1999).

\bibitem{christodoulou1986problem}
D.~Christodoulou, 
  Comm. Math. Phys. {\bf 105}, 337 (1986).

\bibitem{PhysRevLett.70.9}
M.~W. Choptuik, 
  Phys. Rev. Lett. {\bf 70}, 9
  (1993).

\bibitem{gundlach2003critical}
C.~Gundlach, 
Phys. Rep. {\bf 376}, 339 (2003).

\bibitem{Bizon:2011gg}
P.~Bizon and A.~Rostworowski, 
  Phys. Rev. Lett. {\bf 107}, 031102 (2011).

\bibitem{Maliborski:2013via} 
  M.~Maliborski and A.~Rostworowski,
  Int.\ J.\ Mod.\ Phys.\ A {\bf 28}, 1340020 (2013)

\bibitem{Deppe:2016gur} 
  N.~Deppe,
  arXiv:1606.02712 [gr-qc] (2016).

\bibitem{carles2002remarks}
R.~Carles, 
``Remarks on nonlinear Schr{\"o}dinger equations with harmonic
  potential,'' in {\em Annales Henri Poincar{\'e}}, vol.~3, pp.~757--772,
  Springer, 2002. 

\bibitem{jao2016energy}
C.~Jao, 
  Comm.  Partial Differential Equations, {\bf 41}, 
  79, (2016).

\bibitem{hani2015asymptotic}
 Z.~Hani and L.~Thomann, 
   Comm.  Pure  Appl. Math., 2015, Wiley Online Library.

\bibitem{Basu:2014sia}
P.~Basu, C.~Krishnan, and A.~Saurabh, 
   Int. J. Mod. Phys. A {\bf 30},
  1550128 (2015).



\bibitem{Okawa:2013jba}
H.~Okawa, V.~Cardoso, and P.~Pani, 
   Phys. Rev. D {\bf 89}, 041502 (2014).

\bibitem{buchel2013boson}
A.~Buchel, S.~L. Liebling, and L.~Lehner, 
   Phys. Rev. D {\bf 87}, 123006 (2013).

\bibitem{daSilva:2016nah}
E.~da~Silva, E.~Lopez, J.~Mas, and A.~Serantes, 
 J. High Energy Phys {\bf 06}, 172 (2016).

\bibitem{arias2016stability}
R.~Arias, J.~Mas, and A.~Serantes, 
 J. High Energy Phys. {\bf 09}, 024 (2016).

\bibitem{deppe2015stability}
N.~Deppe, A.~Kolly, A.~Frey, and G.~Kunstatter, 
  Phys. Rev. Lett. {\bf 114}, 071102 (2015).

\bibitem{Deppe:2016dcr}
N.~Deppe, A.~Kolly, A.~R. Frey, and G.~Kunstatter, 
 J. High Energy Phys {\bf 10}, 087 (2016).

\bibitem{barcelo2005analogue} C.~Barcel\'o, S.~Liberati,
and M.~Visser,
Living Rev. Rel. {\bf 8}, 214 (2005).

\bibitem{Philbin1367} T.G. Philbin,
C. Kuklewicz, S. Robertson, S. Hill, F. K\"onig, and U. Leonhardt,
Science, {\bf 319}, 1367-1370 (2008).

\bibitem{Hossenfelder1} S. Hossenfelder,
Phys. Rev. D {\bf 91}, 124064 (2015).

\bibitem{Hossenfelder2} S. Hossenfelder,
Phys. Lett. B {\bf 752}, 13-17 (2016).


\bibitem{PhysRevLett.106.115303}
G.~Krstulovic and M.~Brachet, 
  Phys. Rev. Lett. {\bf 106},
  115303, (2011).

\bibitem{shukla2013turbulence}
V.~Shukla, M.~Brachet, and R.~Pandit, 
   New J. Phys.
  {\bf 15}, 113025 (2013).


\bibitem{Balasubramanian:2014cja} 
  V.~Balasubramanian, A.~Buchel, S.~R.~Green, L.~Lehner and S.~L.~Liebling,
  Phys.\ Rev.\ Lett.\  {\bf 113}, no. 7, 071601 (2014).

\bibitem{Craps:2014vaa} 
  B.~Craps, O.~Evnin and J.~Vanhoof,
  J. High Energy Phys. {\bf 10}, 048 (2014).





\bibitem{Craps:2014jwa}
B.~Craps, O.~Evnin, and J.~Vanhoof, 
 J. High Energy Phys. {\bf 01}, 108 (2015).

\bibitem{Buchel:2014xwa}
A.~Buchel, S.~R. Green, L.~Lehner, and S.~L. Liebling, 
   Phys. Rev. D {\bf 91}, 064026 (2015).

\bibitem{chin2010feshbach}
C.~Chin, R.~Grimm, P.~Julienne, and E.~Tiesinga, 
  Rev. Mod. Phys. {\bf 82}, 1225 (2010).

\bibitem{gorlitz2001realization}
A.~G{\"o}rlitz, J.M.~Vogels, A.E.~Leanhardt, C.~Raman, T.L.~Gustavson, J.R.~Abo-Shaeer,
  A.P.~Chikkatur, S.~Gupta, S.~Inouye, T.~Rosenband, and W.~Ketterle
   Phys. Rev. Lett.
  {\bf 87}, 130402 (2001).

\bibitem{PhysRevLett.13.479}
R.~Y. Chiao, E.~Garmire, and C.~H. Townes, 
   Phys. Rev. Lett. {\bf 13}, 479 (1964).

\bibitem{PhysRevLett.90.203902}
K.~D. Moll, A.~L. Gaeta, and G.~Fibich, 
  Phys. Rev. Lett. {\bf 90},
  203902,  (2003).

\bibitem{herring2008radially}
G.~Herring, L.D.~Carr, R.~Carretero-Gonz{\'a}lez, P.G.~Kevrekidis, and
  D.J.~Frantzeskakis, 
   Phys. Rev. A {\bf 77}, 
  023625 (2008).

\bibitem{PhysRevA.51.4704}
P.~A. Ruprecht, M.~J. Holland, K.~Burnett, and M.~Edwards, 
   Phys. Rev. A {\bf 51}, 4704
  (1995).

\bibitem{PhysRevE.92.013201}
K.~Mallory and R.~A. Van~Gorder, 
   Phys. Rev. E, {\bf 92}, 013201
  (2015).

\bibitem{cardoso2016collapsing}
V.~Cardoso and J.~V. Rocha, 
 Phys. Rev. D, {\bf 93}, 084034 (2016).

\bibitem{Maliborski:2014rma}
M.~Maliborski and A.~Rostworowski, 
   Phys. Rev. D {\bf 89}, 124006 (2014).

\bibitem{Cai:2015jbs}
R.-G. Cai, L.-W. Ji, and R.-Q. Yang, 
  Commun. Theor. Phys. {\bf 65}, 
  329 (2016).


  
  \bibitem{Craps:2015iia} 
  B.~Craps, O.~Evnin and J.~Vanhoof,
  J. High Energy Phys. {\bf 10}, 079 (2015).
  
  
  \bibitem{carles2010energy}
 R.~ Carles,  and E.~Faou,
 arXiv:1010.5173 (2010).
  
  \bibitem{turbu1}
  R. Numasato, M. Tsubota, and V.S.~L'vov,
  Phys. Rev. A 81, 063630  (2010).
  
\bibitem{turbu2}
 A.S.~Bradley,  and B.P.~Anderson,
Phys. Rev. X 2 , 041001 (2012).
  
\bibitem{Tsatsos2016}  
M. C. Tsatsos,  P. E.S. Tavares, A. Cidrim, A. R. Fritsch, M. A. Caracanhas, F. E. A. dos Santos, C. F. Barenghi, and V. S. Bagnato,
Phys. Rep. 622 1-52 (2016) 

\bibitem{Evnin:2016mjx} 
  O.~Evnin and P.~Jai-akson,
  J. High Energy Phys. {\bf 04}, 054 (2016).
  
\bibitem{Digital} NIST Digital Library of Mathematical Functions. http://dlmf.nist.gov/18.15, 
Release 1.0.13 of 2016-09-16. F. W. J. Olver, A. B. Olde Daalhuis, D. W. Lozier, B. I. 
Schneider, R. F. Boisvert, C. W. Clark, B. R. Miller, and B. V. Saunders, eds.


\end{thebibliography}
\end{document}